\pdfoutput=1
\documentclass[12pt,a4paper]{article}

\usepackage{ifthen}
\newboolean{pdflatex}
\setboolean{pdflatex}{true}

\newboolean{articletitles}
\setboolean{articletitles}{true}

\newboolean{uprightparticles}
\setboolean{uprightparticles}{false}

\usepackage{microtype}
\usepackage{lineno}
\usepackage{xspace}
\usepackage{graphicx}
\usepackage{color}
\usepackage{colortbl}
\graphicspath{{./}}
\usepackage{amsmath}
\usepackage{amssymb}
\usepackage{amsfonts}
\usepackage{upgreek}

\newcommand*\patchAmsMathEnvironmentForLineno[1]{%
\expandafter\let\csname old#1\expandafter\endcsname\csname #1\endcsname
\expandafter\let\csname oldend#1\expandafter\endcsname\csname
end#1\endcsname
 \renewenvironment{#1}%
   {\linenomath\csname old#1\endcsname}%
   {\csname oldend#1\endcsname\endlinenomath}%
}
\newcommand*\patchBothAmsMathEnvironmentsForLineno[1]{%
  \patchAmsMathEnvironmentForLineno{#1}%
  \patchAmsMathEnvironmentForLineno{#1*}%
}
\AtBeginDocument{%
\patchBothAmsMathEnvironmentsForLineno{equation}%
\patchBothAmsMathEnvironmentsForLineno{align}%
\patchBothAmsMathEnvironmentsForLineno{flalign}%
\patchBothAmsMathEnvironmentsForLineno{alignat}%
\patchBothAmsMathEnvironmentsForLineno{gather}%
\patchBothAmsMathEnvironmentsForLineno{multline}%
}

\usepackage{hyperref}
\usepackage[all]{hypcap}

\def\lhcb {\mbox{LHCb}\xspace}
\def\ux85 {\mbox{UX85}\xspace}

\def\lhc    {\mbox{LHC}\xspace}

\def\cdf    {\mbox{CDF}\xspace}

\ifthenelse{\boolean{uprightparticles}}%
{

 \def\Pmu         {\ensuremath{\upmu}\xspace}

 \def\Ppi         {\ensuremath{\uppi}\xspace}

 \def\Ppsi        {\ensuremath{\uppsi}\xspace}

 \def\PDelta      {\ensuremath{\Delta}\xspace}                 
 \def\PXi      {\ensuremath{\Xi}\xspace}                 
 \def\PLambda      {\ensuremath{\Lambda}\xspace}                 
 \def\PSigma      {\ensuremath{\Sigma}\xspace}                 
 \def\POmega      {\ensuremath{\Omega}\xspace}                 
 \def\PUpsilon      {\ensuremath{\Upsilon}\xspace}                 
 
 \def\PB      {\ensuremath{\mathrm{B}}\xspace}                 
                  
 \def\PD      {\ensuremath{\mathrm{D}}\xspace}

 \def\PJ      {\ensuremath{\mathrm{J}}\xspace}                 
 \def\PK      {\ensuremath{\mathrm{K}}\xspace}

 \def\Pb      {\ensuremath{\mathrm{b}}\xspace}                 
 \def\Pc      {\ensuremath{\mathrm{c}}\xspace}

 \def\Pi      {\ensuremath{\mathrm{i}}\xspace}

 \def\Ps      {\ensuremath{\mathrm{s}}\xspace}

}
{

 \def\Pmu         {\ensuremath{\mu}\xspace}

 \def\Ppi         {\ensuremath{\pi}\xspace}

 \def\Ppsi        {\ensuremath{\psi}\xspace}                 
                  
 \mathchardef\PDelta="7101
 \mathchardef\PXi="7104
 \mathchardef\PLambda="7103
 \mathchardef\PSigma="7106
 \mathchardef\POmega="710A
 \mathchardef\PUpsilon="7107
                  
 \def\PB      {\ensuremath{B}\xspace}                 
                  
 \def\PD      {\ensuremath{D}\xspace}

 \def\PJ      {\ensuremath{J}\xspace}                 
 \def\PK      {\ensuremath{K}\xspace}

 \def\Pb      {\ensuremath{b}\xspace}                 
 \def\Pc      {\ensuremath{c}\xspace}

 \def\Pi      {\ensuremath{i}\xspace}

 \def\Ps      {\ensuremath{s}\xspace}

}

\def\mumu       {\ensuremath{\Pmu^+\Pmu^-}\xspace}

\def\squark    {\ensuremath{\Ps}\xspace}

\def\cquark    {\ensuremath{\Pc}\xspace}
\def\cquarkbar {\ensuremath{\overline \cquark}\xspace}

\def\bquark    {\ensuremath{\Pb}\xspace}

\def\pion  {\ensuremath{\Ppi}\xspace}

\def\pip   {\ensuremath{\pion^+}\xspace}

\def\pipi  {\ensuremath{\pion^+\pion^-}\xspace}

\def\kaon  {\ensuremath{\PK}\xspace}
  \def\Kbar  {\kern 0.2em\overline{\kern -0.2em \PK}{}\xspace}

\def\Kz    {\ensuremath{\kaon^0}\xspace}
\def\Kzb   {\ensuremath{\Kbar^0}\xspace}
\def\KzKzb {\ensuremath{\Kz \kern -0.16em \Kzb}\xspace}
\def\Kp    {\ensuremath{\kaon^+}\xspace}
\def\Km    {\ensuremath{\kaon^-}\xspace}

\def\KpKm  {\ensuremath{\Kp \kern -0.16em \Km}\xspace}

\def\Kstarzb {\ensuremath{\Kbar^{*0}}\xspace}

  \def\Dbar    {\kern 0.2em\overline{\kern -0.2em \PD}{}\xspace}
\def\D       {\ensuremath{\PD}\xspace}

\def\Dz      {\ensuremath{\D^0}\xspace}
\def\Dzb     {\ensuremath{\Dbar^0}\xspace}
\def\DzDzb   {\ensuremath{\Dz {\kern -0.16em \Dzb}}\xspace}
\def\Dp      {\ensuremath{\D^+}\xspace}
\def\Dm      {\ensuremath{\D^-}\xspace}

\def\DpDm    {\ensuremath{\Dp {\kern -0.16em \Dm}}\xspace}

\def\B       {\ensuremath{\PB}\xspace}
  \def\Bbar    {\kern 0.18em\overline{\kern -0.18em \PB}{}\xspace}

\def\Bs      {\ensuremath{\B^0_\squark}\xspace}
\def\Bsb     {\ensuremath{\Bbar^0_\squark}\xspace}
\def\Bdb     {\ensuremath{\Bbar^0}\xspace}

\def\jpsi     {\ensuremath{{\PJ\mskip -3mu/\mskip -2mu\Ppsi\mskip 2mu}}\xspace}

  \def\Y#1S{\ensuremath{\PUpsilon{(#1S)}}\xspace}

\def\Lbar {\ensuremath{\kern 0.1em\overline{\kern -0.1em\PLambda}}\xspace}

\newcommand{\decay}[2]{\ensuremath{#1\!\to #2}\xspace}         

\def\to                 {\ensuremath{\rightarrow}\xspace}

\def\CP                {\ensuremath{C\!P}\xspace}

\newcommand{\phis}{\ensuremath{\phi_{\squark}}\xspace}

\def\BdbToJPsiKst {\decay{\Bdb}{\jpsi\Kstarzb}}

\def\AT#1     {\ensuremath{A_{\mathrm{T}}^{#1}}\xspace}           

\def\C#1      {\ensuremath{\mathcal{C}_{#1}}\xspace}                       
\def\Cp#1     {\ensuremath{\mathcal{C}_{#1}^{'}}\xspace}                    
\def\Ceff#1   {\ensuremath{\mathcal{C}_{#1}^{\mathrm{(eff)}}}\xspace}        
\def\Cpeff#1  {\ensuremath{\mathcal{C}_{#1}^{'\mathrm{(eff)}}}\xspace}       
\def\Ope#1    {\ensuremath{\mathcal{O}_{#1}}\xspace}                       
\def\Opep#1   {\ensuremath{\mathcal{O}_{#1}^{'}}\xspace}                    

\newcommand{\tev}{\ensuremath{\mathrm{\,Te\kern -0.1em V}}\xspace}
\newcommand{\gev}{\ensuremath{\mathrm{\,Ge\kern -0.1em V}}\xspace}
\newcommand{\mev}{\ensuremath{\mathrm{\,Me\kern -0.1em V}}\xspace}
\newcommand{\kev}{\ensuremath{\mathrm{\,ke\kern -0.1em V}}\xspace}
\newcommand{\ev}{\ensuremath{\mathrm{\,e\kern -0.1em V}}\xspace}
\newcommand{\gevc}{\ensuremath{{\mathrm{\,Ge\kern -0.1em V\!/}c}}\xspace}
\newcommand{\mevc}{\ensuremath{{\mathrm{\,Me\kern -0.1em V\!/}c}}\xspace}
\newcommand{\gevcc}{\ensuremath{{\mathrm{\,Ge\kern -0.1em V\!/}c^2}}\xspace}
\newcommand{\gevgevcccc}{\ensuremath{{\mathrm{\,Ge\kern -0.1em V^2\!/}c^4}}\xspace}
\newcommand{\mevcc}{\ensuremath{{\mathrm{\,Me\kern -0.1em V\!/}c^2}}\xspace}

\def\gsim{{~\raise.15em\hbox{$>$}\kern-.85em
          \lower.35em\hbox{$\sim$}~}\xspace}
\def\lsim{{~\raise.15em\hbox{$<$}\kern-.85em
          \lower.35em\hbox{$\sim$}~}\xspace}

\def\evtgen     {\mbox{\textsc{EvtGen}}\xspace}
\def\pythia     {\mbox{\textsc{Pythia}}\xspace}

\def\geant      {\mbox{\textsc{Geant4}}\xspace}

\def\gauss      {\mbox{\textsc{Gauss}}\xspace}

\def\tell1  {TELL1\xspace}
\def\ukl1   {UKL1\xspace}

\def\BsbToJPsifzero {\decay{\Bsb}{\jpsi f_0(980)}}

\def\fzero {\ensuremath{f_0(980)}\xspace}
\def\fz {\ensuremath{f_0}\xspace}

\usepackage{cite}
\usepackage{mciteplus}

\begin{document}

\renewcommand{\thefootnote}{\fnsymbol{footnote}}
\setcounter{footnote}{1}

\begin{titlepage}
\pagenumbering{roman}

\vspace*{-1.5cm}
\centerline{\large EUROPEAN ORGANIZATION FOR NUCLEAR RESEARCH (CERN)}
\vspace*{1.5cm}
\hspace*{-0.5cm}
\begin{tabular*}{\linewidth}{lc@{\extracolsep{\fill}}r}
\ifthenelse{\boolean{pdflatex}}
{\vspace*{-2.7cm}\mbox{\!\!\!\includegraphics[width=.14\textwidth]{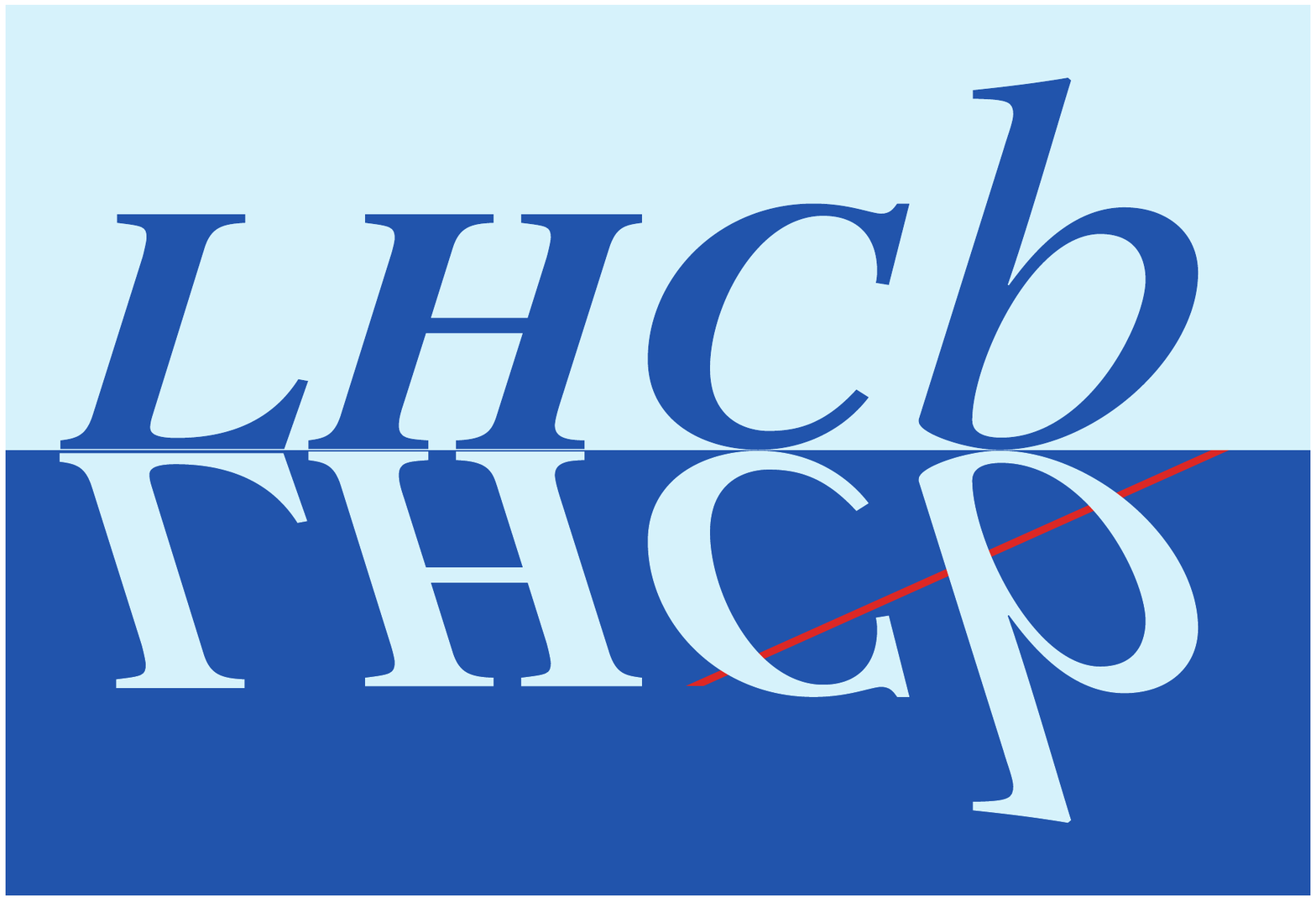}} & &}%
{\vspace*{-1.2cm}\mbox{\!\!\!\includegraphics[width=.12\textwidth]{lhcb-logo.eps}} & &}%
\\
 & & CERN-PH-EP-2012-180 \\ 
 & & LHCb-PAPER-2012-017 \\ 
 & & 4 July 2012
 \end{tabular*}

\vspace*{4.0cm}

{\bf\boldmath\huge
\begin{center}
Measurement of the \Bsb effective lifetime in the $\jpsi\fzero$ final state
\end{center}
}

\vspace*{1.0cm}

\begin{center}
The LHCb collaboration\footnote{Authors are listed on the following pages.}
\end{center}

\vspace{\fill}

\begin{abstract}
\noindent
The effective lifetime of the \Bsb meson in the decay mode \BsbToJPsifzero is measured using 1.0\,fb$^{-1}$ of data collected in $pp$ collisions at $\sqrt{s} = 7$\,TeV with the LHCb detector. The result is $1.700 \pm 0.040 \pm 0.026 \,\mathrm{ps}$ where the first uncertainty is statistical and the second systematic. As the final state is  \CP-odd, and \CP violation in this mode is measured to be small,  the lifetime measurement can be translated into a measurement of the decay width of the heavy \Bsb mass eigenstate, $\Gamma_{\rm H} = 0.588 \pm 0.014 \pm 0.009 \,\rm{ps}^{-1}$. 
\end{abstract}

\vspace*{1.0cm}

\begin{center}
Submitted to Physics Review Letters
\end{center}

\vspace{\fill}

\end{titlepage}

\centerline{\large\bf LHCb collaboration}
\begin{flushleft}
\small
R.~Aaij$^{38}$, 
C.~Abellan~Beteta$^{33,n}$, 
A.~Adametz$^{11}$, 
B.~Adeva$^{34}$, 
M.~Adinolfi$^{43}$, 
C.~Adrover$^{6}$, 
A.~Affolder$^{49}$, 
Z.~Ajaltouni$^{5}$, 
J.~Albrecht$^{35}$, 
F.~Alessio$^{35}$
% ...

{\footnotesize \it
$ ^{1}$Centro Brasileiro de Pesquisas F\'{i}sicas (CBPF), Rio de Janeiro, Brazil\\
% ...
$ ^{5}$Clermont Universit\'{e}, Universit\'{e} Blaise Pascal, CNRS/IN2P3, LPC, Clermont-Ferrand, France\\
$ ^{6}$CPPM, Aix-Marseille Universit\'{e}, CNRS/IN2P3, Marseille, France\\
% ...
$ ^{11}$Physikalisches Institut, Ruprecht-Karls-Universit\"{a}t Heidelberg, Heidelberg, Germany\\
% ...
$ ^{33}$Universitat de Barcelona, Barcelona, Spain\\
$ ^{34}$Universidad de Santiago de Compostela, Santiago de Compostela, Spain\\
$ ^{35}$European Organization for Nuclear Research (CERN), Geneva, Switzerland\\
% ...
$ ^{38}$Nikhef National Institute for Subatomic Physics, Amsterdam, The Netherlands\\
% ...
$ ^{43}$H.H. Wills Physics Laboratory, University of Bristol, Bristol, United Kingdom\\
% ...
$ ^{49}$Oliver Lodge Laboratory, University of Liverpool, Liverpool, United Kingdom\\
% ...
$ ^{n}$LIFAELS, La Salle, Universitat Ramon Llull, Barcelona, Spain\\
% ...
}
\end{flushleft}

\cleardoublepage

\renewcommand{\thefootnote}{\arabic{footnote}}
\setcounter{footnote}{0}

\pagestyle{plain}
\setcounter{page}{1}
\pagenumbering{arabic}

The decay \BsbToJPsifzero, \decay{\fzero}{\pipi}, discovered by \lhcb~\cite{LHCb-PAPER-2011-002} at close to the predicted rate~\cite{Stone:2008ak}, is important for  \CP violation~\cite{LHCb-PAPER-2011-031} and lifetime studies.  In this Letter, we make a precise determination of the lifetime. The $\jpsi\fzero$ final state is \CP-odd, and in the absence of \CP violation, can be produced only by the decay of the heavy ($\rm{H}$), and not by the light ($\rm{L}$), \Bsb mass eigenstate~\cite{Aaij:2012eq}. As the measured \CP violation in this final state is small \cite{LHCb-CONF-2012-002}, a measurement of the effective lifetime, $\tau_{\jpsi\fz}$, can be translated into a measurement of the decay width, $\Gamma_{\rm H}$. This helps to determine the decay width difference, $\Delta\Gamma_{\rm s}=\Gamma_{\rm L}-\Gamma_{\rm H}$, a number of considerable interest for studies of physics beyond the Standard Model (SM)~\cite{Freytsis:2012ja,*Lenz:2012az,*Bobeth:2011st}. Furthermore, this measurement can be used as a constraint in the fit that determines the mixing-induced \CP-violating phase in \Bsb decays, $\phis$, using the $\jpsi\phi$ and $\jpsi f_0(980)$ final states, and thus improve the accuracy of the \phis determination \cite{Fleischer:2011cw,LHCb-CONF-2012-002}. In the SM, if sub-leading penguin contributions are neglected, $\phis = -2\arg\left[ \frac{V_{ts}^{\phantom{*}}V_{tb}^*}{V_{cs}^{\phantom{*}}V_{cb}^*}\right]$, where the $V_{ij}$ are the Cabibbo-Kobayashi-Maskawa matrix elements, which has a value of $-0.036\,^{+0.0016}_{-0.0015}$\,rad \cite{Charles:2011va}. Note that the LHCb measurement of $\phis$~\cite{LHCb-CONF-2012-002} corresponds to a limit on $\cos\phis$ greater than 0.99 at 95\% confidence level, consistent with the SM prediction. 

The decay time evolution for the sum of \Bs and \Bsb decays, via the \decay{\bquark}{\cquark\cquarkbar\squark} tree amplitude, to a \CP-odd final state, $f_-$, is given by~\cite{Nierste:2009wg,*Bigi:2000yz}
\begin{equation}
\Gamma\left(\Bs\to f_-\right) +  \Gamma\left( \Bsb\to f_-\right)=
\frac{ \cal N}{2}e^{-\Gamma_{\rm s} t}\, \Bigg\{e^{\Delta\Gamma_{\rm s} t/2}(1+\cos\phis)+ e^{-\Delta\Gamma_{\rm s} t/2}(1-\cos\phis)\Bigg\}\,,
\label{e:doubleexp}
\end{equation}
where ${\cal N}$ is a time-independent normalisation factor and $\Gamma_{\rm s}$ is the average decay width. We measure the {\it effective lifetime} by describing the decay time distribution with a single exponential function
\begin{equation}
\Gamma\left( \Bs\to f_-\right) + \Gamma\left( \Bsb\to f_-\right) = { \cal N}e^{-t/\tau_{\jpsi\fz}}.
\label{e:singleexp}
\end{equation}
Our procedure involves measuring the lifetime with respect to the well measured \Bdb lifetime, in the decay mode \BdbToJPsiKst, \decay{\Kstarzb}{\Km\pip} (the inclusion of charge conjugate modes is implied throughout this Letter). In this ratio, the systematic uncertainties largely cancel.
 
The data sample consists of 1.0\,fb$^{-1}$ of integrated luminosity collected with the \lhcb detector~\cite{Alves:2008zz} in $pp$ collisions at the \lhc with 7\,\tev centre-of-mass energy. The detector is a single-arm forward spectrometer covering the \mbox{pseudorapidity} range $2<\eta <5$, designed for the study of particles containing \bquark or \cquark quarks. The detector includes a high precision tracking system consisting of a silicon-strip vertex detector surrounding the $pp$ interaction region, a large-area silicon-strip detector located upstream of a dipole magnet and three stations of silicon-strip detectors and straw drift-tubes placed downstream. Charged hadrons are identified using two ring-imaging Cherenkov (RICH) detectors. Muons are identified by a muon system composed of alternating layers of iron and multiwire proportional chambers. The trigger consists of a hardware stage, based on information from the calorimeter and muon systems, followed by a software stage that applies a full event reconstruction. The simulated events used in this analysis are generated using \pythia~6.4~\cite{Sjostrand:2006za} with a specific \lhcb configuration~\cite{LHCb-PROC-2010-056}, where decays of hadronic particles are described by \evtgen~\cite{Lange:2001uf}, and the \lhcb detector simulation~\cite{LHCb-PROC-2011-006} based on \geant~\cite{Allison:2006ve, *Agostinelli:2002hh}.

The selection criteria we use for this analysis are the same as those used to measure $\phis$~in \decay{\Bsb}{\jpsi\pipi} decays~\cite{LHCb-PAPER-2012-006}. Events are triggered by a \decay{\jpsi}{\mumu} decay, requiring two identified muons with opposite charge, transverse momentum greater than 500~MeV (we work in units where $c=\hbar=1$), invariant mass within 120~MeV of the \jpsi mass~\cite{Beringer:2012}, and form a vertex with a fit $\chi^2$ less than 16. $\jpsi\pipi$ candidates are first selected by pairing an opposite sign pion combination with a \jpsi candidate that has a dimuon invariant mass from -48 MeV to +43 MeV from the \jpsi mass~\cite{Beringer:2012}. The pions are required to be identified positively in the RICH detector, have a minimum distance of approach with respect to the primary vertex (impact parameter) of greater than 9 standard deviation significance, have a transverse momentum greater than 250~MeV and fit to a common vertex with the \jpsi with a $\chi^2$ less than 16. Furthermore, the $\jpsi\pipi$ candidate must have a vertex with a fit $\chi^2$ less than~10, flight distance from production to decay vertex greater than 1.5~mm and the angle between the combined momentum vector of the decay products and the vector formed from the positions of the primary and the \Bsb decay vertices (pointing angle) is required to be consistent with zero. Events satisfying this preselection are then further filtered using requirements determined using a Boosted Decision Tree (BDT)~\cite{Roe,*Hocker:2007ht}. The BDT uses nine variables to differentiate signal from background: the identification quality of each muon, the probability that each pion comes from the primary vertex, the transverse momentum of each pion, the \Bsb vertex fit quality, flight distance from production to decay vertex and pointing angle. It is trained with simulated \BsbToJPsifzero signal events and two background samples from data, the first with like-sign pions with $\jpsi\pi^\pm\pi^\pm$ mass within $\pm50$~MeV of the \Bsb mass and the second from the \Bsb upper mass sideband with $\jpsi\pipi$ mass between 200 and 250~MeV above the \Bsb mass. 

\begin{figure}
\begin{center}
    \includegraphics[width=0.495\textwidth]{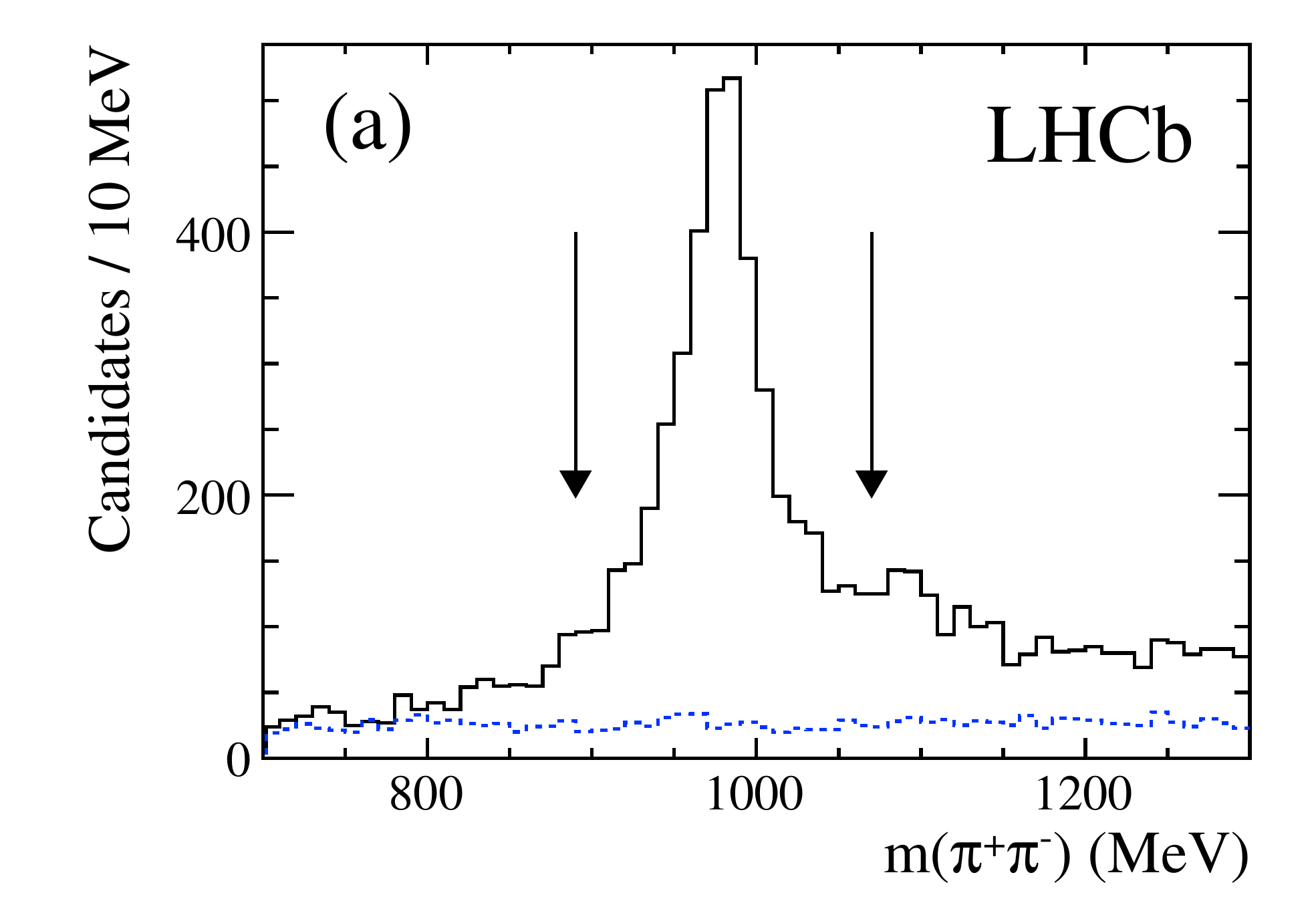}
    \includegraphics[width=0.495\textwidth]{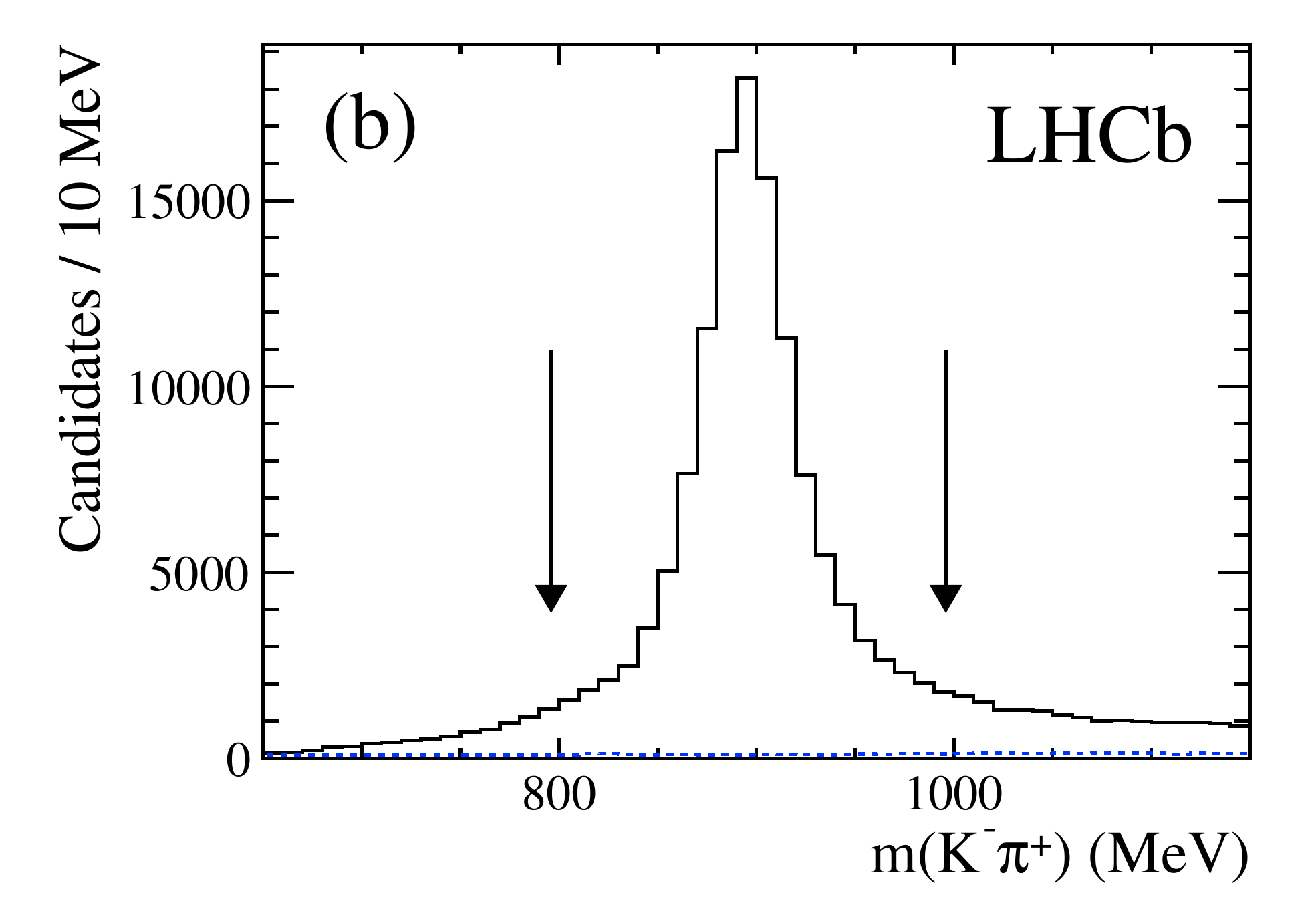}
	\caption{Invariant mass distributions of selected (a) \pipi and (b) $\Km\pip$ combinations (solid histograms) for events within $\pm20$\,MeV of the respective \Bsb and \Bdb mass peaks. Backgrounds (dashed histograms) are determined by fitting the $\jpsi\pipi$ ($\jpsi\Km\pip$) mass in bins of $\pipi$ ($\Km\pip$) mass. Regions between the arrows are used in the subsequent analysis.}
	\label{f:masspi}
\end{center}
\end{figure}

As the effective \BsbToJPsifzero lifetime is measured relative to that of the decay \BdbToJPsiKst, we use the same trigger, preselection and BDT to select $\jpsi\Km\pip$ events, except for the hadron identification that is applied independently of the BDT. The selected \pipi and $\Km\pip$ invariant mass distributions, for candidates with $\jpsi\pipi$ ($\jpsi\Km\pip$) mass within $\pm20$\,MeV of the respective \B mass peaks are shown in Fig.~\ref{f:masspi}. The background distributions shown are determined by fitting the $\jpsi\pipi$ ($\jpsi\Km\pip$) mass distribution in bins of $\pipi$ ($\Km\pip$) mass. Further selections of $\pm90$\,\mev around the \fzero mass and $\pm 100$\,\mev around the \Kstarzb mass are applied. The \fzero selection results in a \BsbToJPsifzero  sample that is greater than 99.4\% \CP-odd at 95\% confidence level~\cite{LHCb-PAPER-2012-005}. 

The analysis exploits the fact that the kinematic properties of the \BsbToJPsifzero decay are very similar to those of the \BdbToJPsiKst decay. We can select \B mesons in either channel using identical kinematic constraints and hence the decay time acceptance introduced by the trigger, reconstruction and selection requirements should almost cancel in the ratio of the decay time distributions. Therefore, we can determine the \BsbToJPsifzero lifetime, $\tau_{\jpsi\fz}$, relative to the \BdbToJPsiKst lifetime, $\tau_{\jpsi\Kstarzb}$, from the variation of the ratio of the \B meson yields with decay time
\begin{equation}
R(t)= R(0)e^{-t(1/\tau_{\jpsi\fz} - 1/\tau_{\jpsi\Kstarzb})} = R(0)e^{-t\Delta_{\jpsi\fz}} \,\mathrm{,}
\label{e:delta}
\end{equation}
where the width difference $\Delta_{\jpsi\fz} = 1/\tau_{\jpsi\fz} - 1/\tau_{\jpsi\Kstarzb}$.

\begin{figure}
\begin{center}
	    \includegraphics[width=0.6\textwidth]{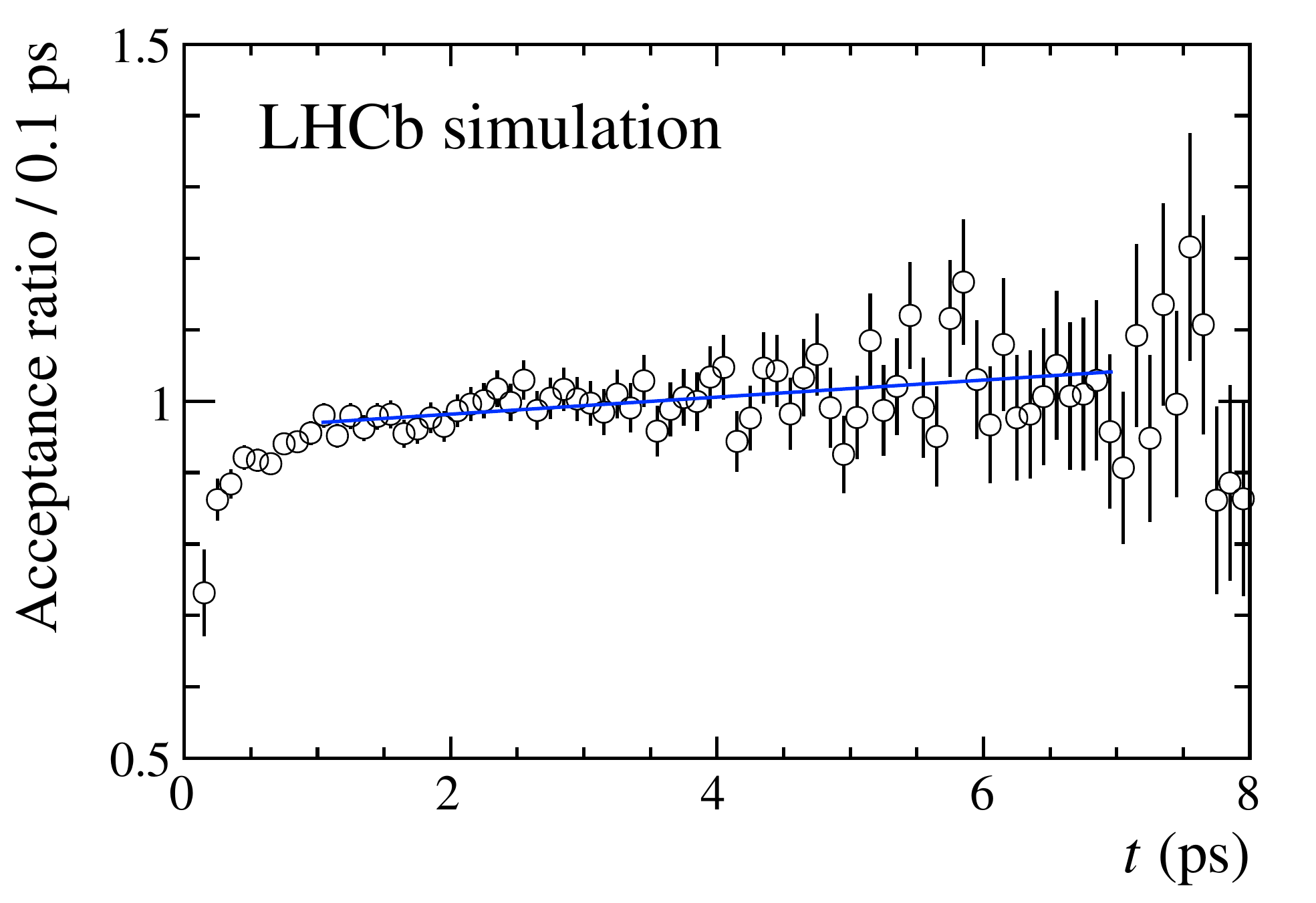}
	\caption{Ratio of decay time acceptances between \BsbToJPsifzero and \BdbToJPsiKst decays obtained from simulation. The solid (blue) line shows the result of a linear fit.}
		\label{f:acceptance}
\end{center}
\end{figure}

We test the cancellation of acceptance effects using simulated \BsbToJPsifzero and \BdbToJPsiKst events. Both the acceptances themselves and also the ratio exhibit the same behaviour. Due to the selection requirements, they are equal to 0 at $t = 0$, after which there is a sharp increase, followed by a slow variation for $t$ greater then 1\,ps. Based on this, we only use events with  $t$ greater than 1\,ps in the analysis. To good approximation, the acceptance ratio is linear between 1 and 7\,ps, with a slope of $a = 0.0125 \pm 0.0036$\,ps$^{-1}$ (see Fig.~\ref{f:acceptance}). We use this slope as a correction to Eq. \ref{e:delta} when fitting the measured decay time ratio
\begin{equation}
R(t)  = R_0(1+at)e^{-t\Delta_{\jpsi\fz}}.
\label{e:deltacorrected}
\end{equation}

Differences between the decay time resolutions of the decay modes could affect the decay time ratio. To measure the decay time resolution, we use prompt events containing a \jpsi meson. Such events are found using a dimuon trigger, plus two opposite-charged tracks with similar selection criteria as for $\jpsi\pipi$ ($\jpsi\Km\pip$) events, apart from any decay time biasing requirements such as impact parameters and \B flight distance, additionally including that the $\jpsi\pipi$ ($\jpsi\Km\pip$) mass be within $\pm20$\,\mev of the \Bsb (\Bdb) mass. To describe the decay time distribution of these events, we use a triple Gaussian function with a common mean, and two long lived components, modelled by exponential functions convolved with the triple Gaussian function. The events are dominated by zero lifetime background with the long lived components comprising less than 5\% of the events. We find the average effective decay time resolution for \BsbToJPsifzero and \BdbToJPsiKst decays to be $41.0 \pm 0.9$\,fs and $44.1 \pm 0.2$\,fs respectively, where the uncertainties are statistical only. This difference was found not to bias the decay time ratio using simulated experiments.

In order to determine the \BsbToJPsifzero lifetime, we determine the yield of \B mesons for both decay modes using unbinned maximum likelihood fits to the \B mass distributions in 15 bins of decay time of equal width between 1 and 7\,ps. We perform a $\chi^2$ fit to the ratio of the yields as a function of decay time and determine the relative lifetime according to Eq.~\ref{e:deltacorrected}. We obtain the signal and peaking background shape parameters by fitting the time-integrated dataset. In each decay time bin, we use these shapes and determine the combinatorial background parameters from the upper mass sidebands, $5450 < m(\jpsi\fz) <  5600$\,MeV and $5450 < m(\jpsi\Kstarzb) < 5550$\,MeV. With this approach, the combinatorial backgrounds are re-evaluated in each bin and we make no assumptions on the shape of the  background decay time distributions. This method was tested with high statistics simulated experiments and found to be unbiased. 

\begin{figure}
\begin{center}
    \includegraphics[width=0.495\textwidth]{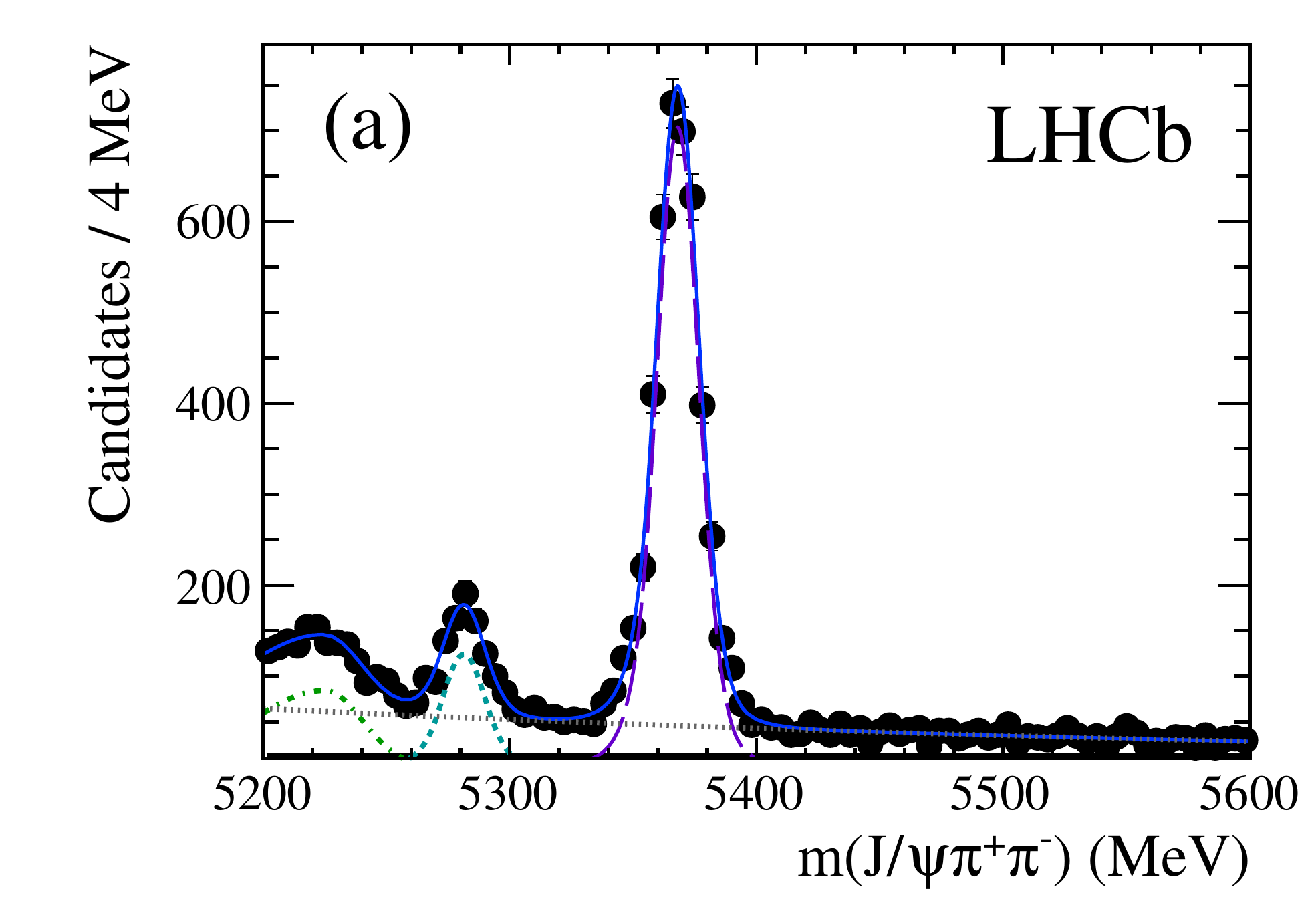}
    \includegraphics[width=0.495\textwidth]{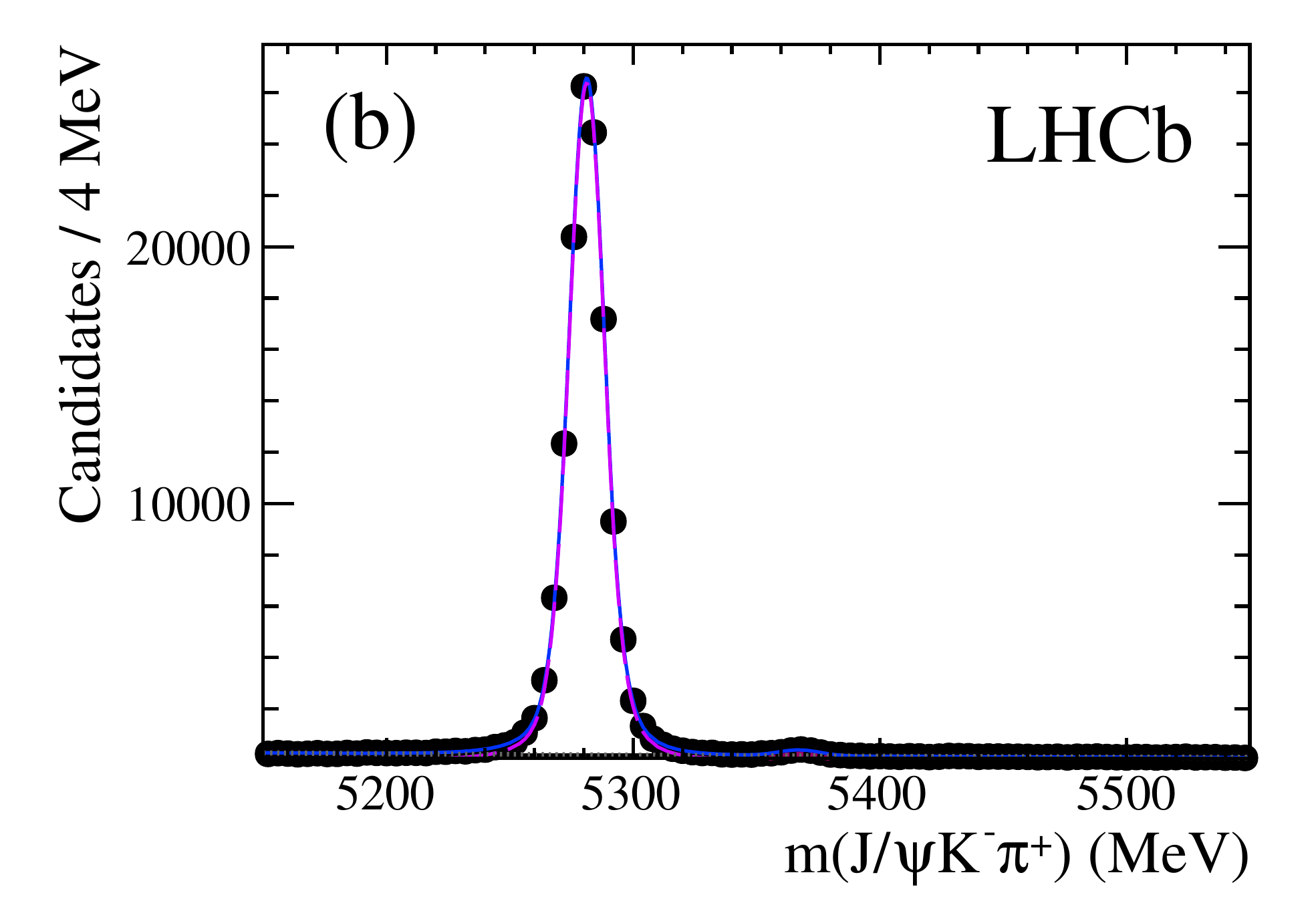}
	\caption{Invariant mass distributions of selected (a) $\jpsi\pipi$ and (b) $\jpsi\Km\pip$ candidates. The solid (blue) curves show the total fits, the long dashed (purple) curves show the respective \BsbToJPsifzero and \BdbToJPsiKst signals, and the dotted (gray) curve shows the combinatorial background. In (a) the short dashed (blue-green) curve shows the \decay{\Bdb}{\jpsi\pipi} background and the dash dotted (green) curve shows the \decay{\Bdb}{\jpsi}{\Km\pip} reflection. In (b) the short dashed (red) curve near 5370 MeV shows the \decay{\Bsb}{\jpsi\Km\pip} background.}
	\label{f:masspsi}
\end{center}
\end{figure}

\begin{figure}
\begin{center}
    \includegraphics[width=0.495\textwidth]{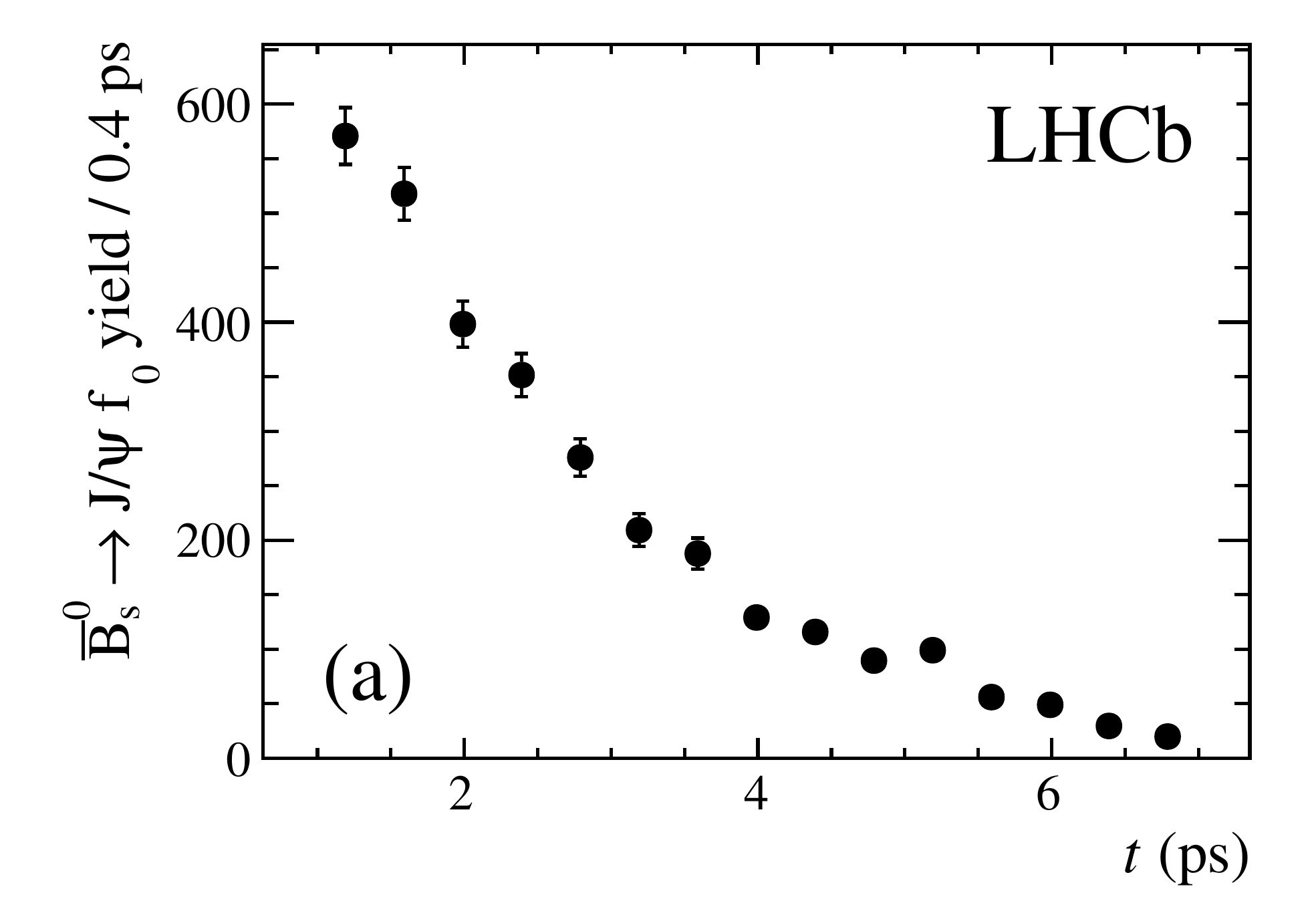}
    \includegraphics[width=0.495\textwidth]{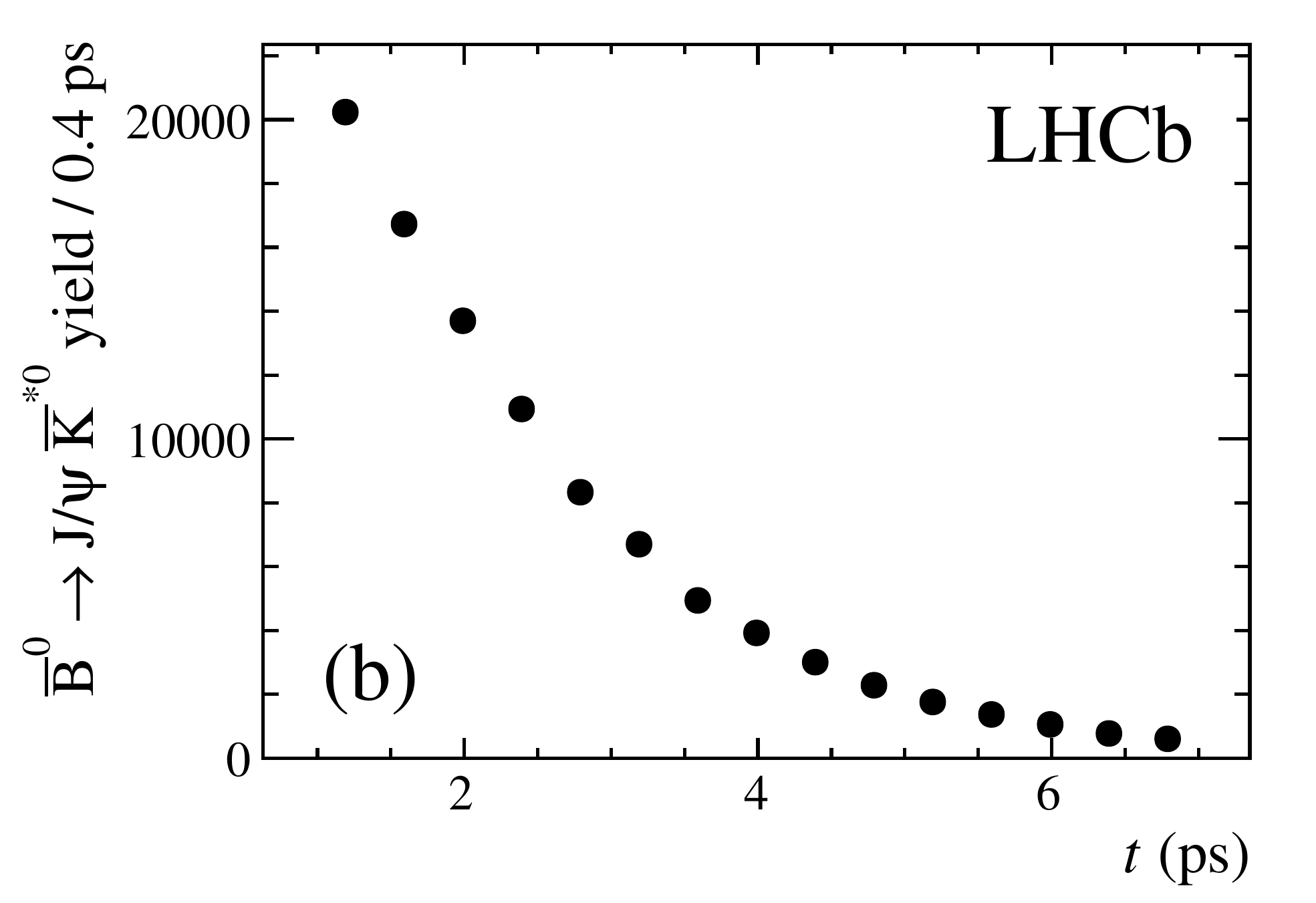}
    \caption{Decay time distributions for (a) \BsbToJPsifzero and (b) \BdbToJPsiKst. In (b) the error bars are smaller than the points.}
    \label{f:lifetime}
\end{center}
\end{figure}

\begin{figure}
\begin{center}
	    \includegraphics[width=0.6\textwidth]{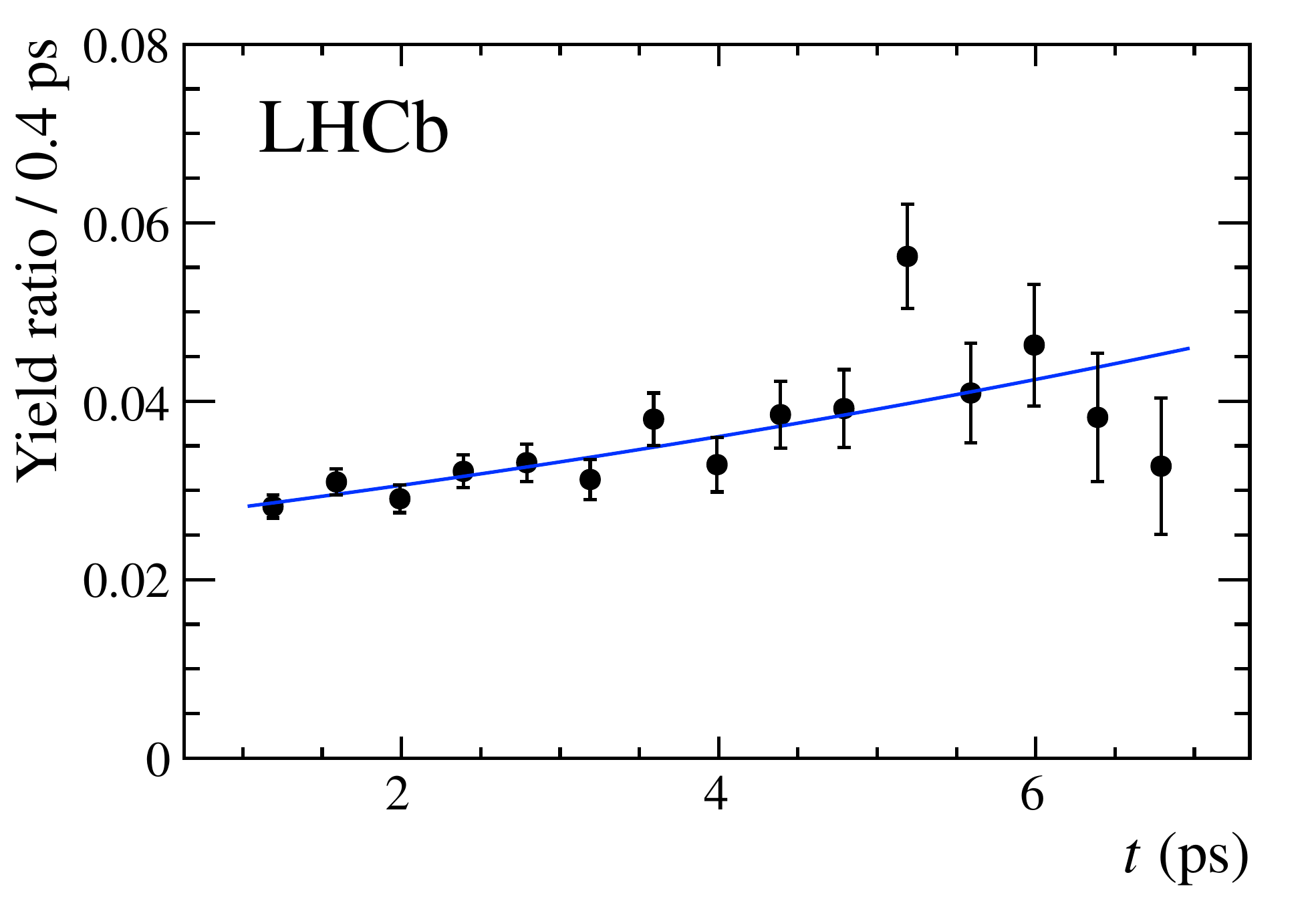}
	\caption{Decay time ratio between \BsbToJPsifzero and \BdbToJPsiKst, and the fit for~$\Delta_{\jpsi\fz}$.}
	\label{f:ratio}
\end{center}
\end{figure}

The time-integrated fits to the $\jpsi\fzero$ and the $\jpsi\Kstarzb$ mass spectra are shown in Fig.~\ref{f:masspsi}. The signal distributions are described by the sum of two Crystal Ball functions \cite{Skwarnicki:1986xj} with common means and resolutions for the Gaussian core, but different parameters describing the tails
\begin{equation}
f(m;\mu,\sigma,n_{l,r},\alpha_{l,r}) =
\begin{cases} 
\left(\frac{n_l}{|\alpha_l|}\right)^{n_l}\cdot\exp\left(\frac{-|\alpha_l|^2}{2}\right)\cdot\left(\frac{n_l}{|\alpha_l|}-|\alpha_l|-\frac{|m-\mu|}{\sigma}\right)^{-n_l}, & \text{if $\frac{m-\mu}{\sigma}\le -\alpha_l$,}\\
\left(\frac{n_r}{|\alpha_r|}\right)^{n_r}\cdot\exp\left(\frac{-|\alpha_r|^2}{2}\right)\cdot\left(\frac{n_r}{|\alpha_r|}-|\alpha_r|-\frac{|m-\mu|}{\sigma}\right)^{-n_r}, & \text{if $\frac{m-\mu}{\sigma}\ge \alpha_r$,} \\
\exp(\frac{-(m-\mu)^2}{2\sigma^2}), &\text{otherwise,}
\end{cases}
\end{equation}
where $\mu$ is the mean and $\sigma$ the width of the core, while $n_{l,r}$ are the exponent of the left and right tails, and $\alpha_{l,r}$ are the left and right transition points between the core and tails. The left hand tail accounts for final state radiation and interactions with matter, while the right hand tail describes non-Gaussian detector effects only seen with increased statistics. The combinatorial backgrounds are described by exponential functions. All parameters are determined from data. There are $4040 \pm 75$ \BsbToJPsifzero and $131\,920 \pm 400$ \BdbToJPsiKst signal decays. The decay time distributions, determined using fits to the invariant mass distributions in bins of decay time as described above, are shown in Fig.~\ref{f:lifetime}. These are made by placing the fitted signal yields at the average \BdbToJPsiKst decay time within the bin rather than at the centre of the decay time bin. This procedure corrects for the exponential decrease of the decay time distributions across the bin. The subsequent decay time ratio distribution is shown in Fig.~\ref{f:ratio}, and the fitted reciprocal lifetime difference is $\Delta_{\jpsi\fz} = -0.070 \pm 0.014 \,\mathrm{ps}^{-1}$, where the uncertainty is statistical only. Taking $\tau_{\jpsi\Kstarzb}$ to be the mean \Bdb lifetime $1.519 \pm 0.007$\,ps~\cite{Beringer:2012}, we determine $\tau_{\jpsi\fz} = 1.700 \pm 0.040\,\mathrm{ps}$.

Sources of systematic uncertainty on the \BsbToJPsifzero lifetime are investigated and listed in Table~\ref{t:systematics}. We first investigate our assumptions about the signal and combinatorial background mass shapes. The relative change of the determined \BsbToJPsifzero lifetime between fits with double Crystal Ball functions and double Gaussian functions for the signal models is 0.001\,ps, and between fits with exponential functions and straight lines for the combinatorial background models is 0.010\,ps. The different particle identification criteria used to select $\Bsb \rightarrow \jpsi\fzero \rightarrow\mumu\pipi$ and $\Bdb \rightarrow \jpsi\Kstarzb \rightarrow\mumu\Km\pip$ decays could affect the acceptance cancellation between the modes. In order to investigate this effect, we loosen and tighten the particle identification selection for the kaon, modifying the \BdbToJPsiKst signal yield by $+2$\% and $-20$\% respectively, and repeat the analysis. The larger difference with respect to the default selection, 0.007\,ps, is assigned as a systematic uncertainty. We also assign half of the relative change between the fit without the acceptance correction and the default fit, 0.018\,ps, as a systematic uncertainty. Potential statistical biases of our method were evaluated with simulated experiments using similar sample sizes to those in data. An average bias of 0.012\,ps is seen and included as a systematic uncertainty. The observed bias vanishes in simulated experiments with large sample sizes. As a cross-check, the analysis is performed with various decay time bin widths and fit ranges, and consistent results are obtained. The possible \CP-even component, limited to be less than 0.6\% at 95\% confidence level~\cite{LHCb-PAPER-2012-005}, introduces a 0.001\,ps systematic uncertainty. Using the PDG value for the \Bdb lifetime~\cite{Beringer:2012} as input requires the propagation of its error as a systematic uncertainty. All the contributions are added in quadrature and yield a total systematic uncertainty on the lifetime of 0.026\,ps (1.5\%). Thus the effective lifetime of the $\jpsi f_0(980)$ final state in \Bsb decays, when describing the decay time distribution as a single exponential is 
\begin{equation}
\tau_{\jpsi\fz}=1.700\pm 0.040\pm 0.026 \, {\rm ps}\,.
\end{equation}

\begin{table}[t]
\begin{center}
\caption{Summary of  systematic uncertainties on the \BsbToJPsifzero effective lifetime.}
\label{t:systematics}
\begin{tabular}{ c  c }
\hline\hline
Source & Uncertainty (ps)\\
\hline
Signal mass shape & $0.001$ \\
Background mass shape & $0.010$ \\
Kaon identification & $0.007$ \\
Acceptance & $0.018$ \\
Statistical bias &  $0.012$ \\
\CP-even component & $0.001$ \\
\Bdb lifetime~\cite{Beringer:2012} & $0.009$ \\
\hline
Sum in quadrature & $0.026$\\
\hline\hline
\end{tabular}
\end{center}
\end{table}

Given that \phis is measured to be small, and the decay is given by a pure \decay{\bquark}{\cquark\cquarkbar\squark} tree amplitude, we may interpret the inverse of the \BsbToJPsifzero effective lifetime as a measurement of $\Gamma_{\rm H}$ with an additional source of systematic uncertainty due to a possible non-zero value of \phis. For $\cos\phis = 0.99$, $\Gamma_{\rm s} = 0.6580$\,ps$^{-1}$ and $\Delta\Gamma_{\rm s} = 0.116$\,ps$^{-1}$~\cite{LHCb-CONF-2012-002}, $\tau_{\jpsi\fz}$ changes by 0.002\,ps. This is added in quadrature to the systematic uncertainties on $\tau_{\jpsi\fz}$ to obtain the final systematic uncertainty on $\Gamma_{\rm H}$.

In summary, the effective lifetime of the \Bsb meson in the \CP-odd $\jpsi\fzero$ final state has been measured with respect to the well measured \Bdb lifetime in the final state $\jpsi\Kstarzb$. The analysis exploits the kinematic similarities between the \BsbToJPsifzero and \BdbToJPsiKst decays to determine an effective lifetime of
\begin{equation*}
	\tau_{\jpsi\fz} = 1.700 \pm 0.040 \pm 0.026 \,\mathrm{ps},
\end{equation*}
corresponding to a width difference of
\begin{equation*}
\Delta_{\jpsi\fz} = -0.070 \pm 0.014 \pm 0.001 \,\mathrm{ps}^{-1},
\end{equation*}
where the uncertainties are statistical and systematic respectively. This result is consistent with, and more precise than, the previous measurement of $1.70\,^{+0.12}_{-0.11}\pm0.03$\,ps from \cdf ~\cite{Aaltonen:2011nk}.
Interpreting this as the lifetime of the heavy \Bsb eigenstate, we obtain
\begin{equation*}
\Gamma_{\rm H} = 0.588 \pm 0.014 \pm 0.009 \,\mathrm{ps}^{-1}.
\end{equation*}
This value of $\Gamma_{\rm H}$ is consistent with the value $0.600 \pm 0.013$\,ps$^{-1}$, calculated from the values of $\Gamma_{\rm s}$ and $\Delta \Gamma_{\rm s}$ in Ref.~\cite{LHCb-CONF-2012-002}.

\section*{Acknowledgements}

\noindent We express our gratitude to our colleagues in the CERN accelerator departments for the excellent performance of the LHC. We thank the technical and administrative staff at CERN and at the LHCb institutes, and acknowledge support from the National Agencies: CAPES, CNPq, FAPERJ and FINEP (Brazil); CERN; NSFC (China); CNRS/IN2P3 (France); BMBF, DFG, HGF and MPG (Germany); SFI (Ireland); INFN (Italy); FOM and NWO (The Netherlands); SCSR (Poland); ANCS (Romania); MinES of Russia and Rosatom (Russia); MICINN, XuntaGal and GENCAT (Spain); SNSF and SER (Switzerland); NAS Ukraine (Ukraine); STFC (United Kingdom); NSF (USA). We also acknowledge the support received from the ERC under FP7 and the Region Auvergne.

\appendix
\addcontentsline{toc}{section}{References}
\ifx\mcitethebibliography\mciteundefinedmacro
\PackageError{LHCb.bst}{mciteplus.sty has not been loaded}
{This bibstyle requires the use of the mciteplus package.}\fi
\providecommand{\href}[2]{#2}

\end{document}